\documentclass[preprint2,twoside]{hwo}
\usepackage[table]{xcolor} 
\usepackage{booktabs} 
\usepackage{lscape}
\usepackage{graphicx}

\input{hwo.h}

\begin{document}

\title{\textbf{\LARGE Advancing the Concept Maturity Level of the Servicing Architecture for Habitable Worlds Observer }}
\author {\textbf{\large Jonathan W. Arenberg$^{1}$ }}
\affil{$^1$\small\it Northrop Grumman Systems Corporation, One Space Park Drive, Redondo Beach, California USA}




\begin{abstract}
 This paper advances the concept maturity level (CML) of the Habitable Worlds Observatory (HWO) servicing architecture. Since servicing has occurred on other missions, this paper argues that the current CML is 2. To advance to CML 3, option spaces must be established for trade studies. We introduce the three space ages and the argument that we are on the cusp of a new revolutionary era.  Servicing is only part of that coming change and the other elements of the future space age are introduced. The challenge of designing a flagship mission such as HWO is discussed. The value proposition for the adoption of a new technology, such as servicing HWO, is established. In the latter portion of the paper it is shown that these elements have promise of being beneficial to HWO and should be included in any trade space. 
  \\
  \\
\end{abstract}

\vspace{2cm}

\section{Introduction}

The Habitable Worlds Observatory (HWO) has been declared serviceable. However, the specific definition of what will and will not be serviced and how that is to be accomplished has not been decided or even studied. In short, the Concept Maturity Level (CML), for the servicing architecture is immature.  CML was developed at the Jet Propulsion Laboratory (JPL) to aid in the mission development process. JPL shared this process in a series of papers in the 2000s. This very useful idea has been widely adopted and is a standard term of art in HWO discussions. \cite{CML09}, \cite{CML10}, \cite{CML13}  

In terms of the CML scale, the maturity of the servicing architecture for HWO is a 2, corresponding to a proof of concept. The proof of concept is Northrop Grumman’s commercial Mission Extension of a previously de-activated communication satellite client satellite, Intelsat IS-901 on February 25, 2020.\cite{MEV}  

To advance to CML 3 we must begin doing trade studies. While that is our goal, the reader will find that there is much work before formal trade studies can begin. This added complexity became obvious as the author realized that the field of space technology is undergoing a period of rapid change, development and advancement, servicing is just the herald of this coming revolution. 

Our discussion will begin with the description of the Space Ages, Old, Middle and New and in particular what characterizes the current Late-Middle Space Age and how New or Neo-Space differs.  The challenges with system design in the current era will be introduced, followed by a discussion of how many elements of the coming space age may mitigate design challenges and open the way for modern system architecture, with integrated servicing for HWO. The paper concludes the next steps needed to continue to develop the HWO concept, so that it is ready for its stated development goal of passing Mission Confirmation Review (MCR) at the end of the current decade.

\section{The Space Ages}

To discern periods of significant technological and societal changes, historians, anthropologists and archeologists have divided human pre-history into ages. The canonical system of dividing this period into three ages, Stone, Bronze and Iron, is credited to Christian Thomsen, a Danish antiquarian, who developed this idea in the early 19th Century.\cite{3ages} The three age division of time was successful, and later in the 19th century John Lubbock subdivided the Stone age into early (paleolithic) and new (neolithic).  The organizations of periods by prevailing technology are very useful in understanding human history and development. We adopt them in this paper to provide a framework for understanding the development of space technology and ultimately how this will drive architecture and design decisions for HWO. 

\subsection{Old Space Age or Paleo-Space Age}

The Old Space Age, Paleo-Space or more compactly, Space 1.0 runs from antiquity to January 27, 1967, the date of the Apollo 1 and the deaths of astronauts, Grisson, Chaffee and White. This period is characterized by the adaptation of hardware produced for other purposes, such as adapting wartime ballistic missiles, such as captured German V-2 rockets in the 1940's and 50's to the missiles that carried the first US astronauts into space and orbit. The prevailing approach was a learn by doing attitude that valued speed and cadence over the deep design  analysis of possible failure modes and effects. 
\\
\\
The ethos for this pioneering era of space technology is \emph{'Just make it work.'}

\subsection{Middle Space Age or Meso-Space}

The demise of the Apollo 1 crew sparked an extensive investigation and review of the Apollo capsule design.   The cause of the deadly fire had many causes but was summarized be board member and astronaut Frank Borman who assigned the cause to \emph{“a failure of imagination”.} \cite{Apollo1} The renewed emphasis on safety and confidence in mission success has laid the foundations of the successes in the American space program crewed and uncrewed since 1968. The central aim of systems design in this period is one of deep understanding of the behavior and response of the system. The profound understanding of these systems allows for designs and operations concepts to prevent these failures.  Systems such as Pioneers 10 and 11, Voyagers 1 and 2 and the Chandra X-ray Observatory the James Webb Space Telescope are examples of such Space 2.0 missions. 
\\

The ethos for the period is certainly, \emph{'Failure is not an option.'}
\\

The Meso-Space age is characterized by the emergence of mission bespoke highly engineered, tested and verified hardware and systems. These systems were designed and manufactured by a limited number of large corporations and nations.  Spacefaring was largely confined to major nation states and an emerging commercial market in communications satellites and systems.

In the late 20th century trends to greater participation in space activities and hardware can be seen in the historical record. In the last decade of the 20th century the first privately funded launcher, Pegasus, had its maiden flight, 5 April 1990. This commercial development signals that spacefaring was no longer only the province of only large nation states.  Smaller entities, countries, corporations and in time even students could make and assemble viable, functioning space hardware for missions of their own design and choosing. 

\subsubsection{Early and Late Meso-Space Ages}

In 2020 as mentioned in the introduction, the Northrop Grumman MEV docked with an unprepared client satellite for an unprecedented life extension mission.\cite{MEV} This mission was an entire commercial effort and serves as the dividing point between the early and late Middle space ages, or Space 2.0 and 2.1. The division between early and late Meso-Space is this servicing mission that forms the basis of our claim of CML for HWO servicing. 

\subsection{New Space Age or Neo-Space} 

The Neo-Space age, or Space 3.0, is nascent. Despite the fact that it has not yet arrived, robotic servicing can be seen as its harbinger, Using some reasonable projections, we can characterize the new age at a high level and define what describes it.

Space 3.0 is characterized by far more participants in space, as mission planners and operators. These missions will be carried out by students, companies and corporations and nation states of all sizes. This expansion of participation is enabled by the expansion of the supplier base, including many new launch opportunities with decreased cost, such as ride-share service to dedicated privately developed very large launchers.  These non-national space farers will bring with them greater risk tolerance, allowing for higher launch cadence and lower cost.  

All of these new players in the space hardware and mission economy create something truly new and revolutionary, namely a viable market, where hardware and systems will be developed for one purpose and then used for another. Purchased commercially with ever-reduced cost and increasing performance.  Examples of this kind of technology adoption of originally a niche novel technology expanding into many markets is the airplane from the early 20th century and Global Positioning Satellite system from the latter part of that century. Originally a very niche market, these technologies are a fundamental part of the fabric of modern life, in the 21st century.

Servicing and logistics will be commonplace and part of mission designs, with appropriate and validated cost modeling that as of 2025, does not exist.

Other elements of Space 3.0 will be an expanded number of designer materials and advanced manufacturing, initial terrestrial and ultimately in space. 

Computing advances will also change the design and operation environment. Even today, artificial intelligence and machine learning are being introduced to the management of large constellations with the promise of increases in cost performance in flight operations. 

Mission technology can be refreshed after launch; servicing allows for replacement of prepared system elements. Science instruments are a prime candidate for refreshment on a servicing mission.   Another obvious technology for servicing is the replacement of aging or failed solar arrays. As has been cited before, life extension, either through adding hardware like the MEV and upcoming MRV missions, already exists commercially but will be widely purveyed in the future by any number of companies.

Communications technologies are advancing, becoming more capable, more data capacity, and less expensive. Furthermore, ground stations to communicate with satellites are becoming widespread as evidenced by CubeSat ground station kits that are widely advertised, as can be confirmed by a simple search.  Optical communications are advancing for governmental and commercial uses, and they are being adopted for up and down links as well as satellite cross links. NASA is investigating commercial systems for lunar communications and navigation services, and many commercial companies have ground stations that can be rented to support space missions.

\section{What Makes Flagship Design Difficult (and Expensive) }

Astrophysics is an observational science, progress in this field has been driven by observation. Since the start of the telescope era in 1609, the field has been paced by the need to make more sensitive systems to make new observations and push the field forward. This is true for ground and space-based observatories. Space based observatories are realized in different size (cost) mission classes, with Flagships being the largest. 

Flagships such as HWO serve as a main engine of scientific advancement for the community. They are part of the community thought and planning from ideation to retirement. At the recent HWO 2025 meeting, it was announced that more than 1,000 people had signed up to be part of the science and mission planning, even noting that the launch is “penciled’" for 2045, 20 years from the time of this writing. 

\subsection{Flagship Development is Intrinsically Hard}

Flagship development is by definition a uniquely challenging endeavor. These endeavors are intrinsically hard, I know this from my personal experience on Chandra and Webb.

The nature of this challenge has been described elsewhere but is only summarized here for the use of this discussion.   The main challenges to Flagship development are:

\begin{itemize}
    \item Need for a large leap in scientific capability,
    \item low tolerance for risk,
    \item Large number of stakeholders,
    \item Low tolerance for risk.
\end{itemize}

These four areas of challenge are endemic to the creation of a large observatory system that serves the broad need of the community and creates new capabilities and observations that drive discovery.

These systems are necessarily complex, technically and programmatically. Put in simple terms, there is no one central problem, that if rectified or changed, would suddenly decrease development time and costs. Put simply, there is no silver bullet that if fired at this problem will “make it easy”. 

\subsection{Self-containment of the flight segment-Tyranny of the Fairing}

Paramount in the design cannon for space observatories in the paleo and meso-space eras is that the entire space faring segment of the system is launched together in a single event. While there have been notable deviations, Hubble is of course famously serviceable and Chandra nee AXAF was also a serviceable mission early in its development. It is noted that Hubble as launched with its inaugural instruments did have to meet the mass and volume requirements of the Space Shuttle. Chandra had the same requirements and adopted a non-serviceable architecture when its budget was reduced by 70\% early in development.\cite{ChandraLL}   The two other great observatories, Compton and Spitzer were not serviceable. 

Under the current design cannon, missions are self-contained and complete, all flight hardware is on board at the time of launch. The systems design must comply with the mass and volumetric constraints of the launch vehicle. 

Moreover, the schedule to get the hardware to launch is regulated by the latest item to be completed. The pressure of building a flagship means that an instrument lagging in development is not likely to be “left on the ground” as that constituency would not have its needs met. It is worth nothing that in the re-architecture of the AXAF (now Chandra) mission, the x-ray calorimeter was deleted from the instrument suite over concerns of cost and schedule, but NASA arranged for its launch on another mission. 

Collectively, the need to get the whole mission in a single fairing and comply with the volume and mass constraints has been called “the tyranny of the fairing”. \cite{Tyranny}

/subsection{Cultural pressure of “Failure is not an option”}

The pressure to meet the ethos of the meso-space age, “failure is not an option” is intense. A Flagship is a big ticket item, expensive in terms of cost, time and talent consumed to create them. Many, like the author, spend large fractions of their careers on these missions and it is natural for us all to want to protect the investments in time and effort we have made. The systems demand novel technologies and architectures interacting in complex ways. The design team must determine in all those combinations that the system only behaves in desired and determinative ways and does not have some failure mechanism that should have been identified. Namely we must avoid the “failure of imagination” that doomed Apollo 1. 

Achieving confidence in behavior of a complex, subtle, novel system, is a special challenge and given the name, the “Lesson of Newness.”\cite{WebbLL}   Trying to drive determinism into the engineering process and performance of the system often results in the use older technologies as their behavior is usually better understood than novel materials and processes. 

Uncertainty in the design process, especially when it comes to interfaces and induced environments, invites a conservative approach to design. This conservative approach usually responds to uncertainty leads to over-design. In short, “uncertainty is expensive”.

\section{HWO as a Space 3.0 era design}

As we have introduced earlier in this discussion, servicing of spacecraft exists and has occurred. For the broad purposes of this discussion, it should be noted that this has happened as a crewed activity as NASA missions and has happened robotically as commercial missions. We could thus conclude that servicing is technically mature, but for HWO it is technically immature, namely we don’t have a clean clear list of what to service and why.  Or more broadly, how to leverage the full range of Space 3.0 attributes to the benefit of the HWO design and mission. 

\subsection{Can The Design Options Available in the Neo-Space Era Help HWO}

Our goal is to establish elements of a trade space that will allow trade studies and advance the CML of servicing from 2 to 3. To find the elements of Space 3.0 that might benefit HWO we need to ask and answer the high-level question, could a given element be of benefit to HWO. If the answer is yes or maybe, then it should be included as factor in a future trade study.   The question of “benefit” is not a simple scalar and so ask a set of questions.

\begin {itemize}
    \item Can the inclusion of (fillin here)  create a science capability that is not otherwise achievable? 
    \item Can the inclusion of (fillin here)  reduce cost, schedule or risk of HWO development?
    \end {itemize}

The answers to an initial set of answers to these questions are shown in Figure 1, the table below. In this table challenges or development needs are listed in the first column. The technologies or impacts of Space 3.0 are placed in the column, with their titles listed on the first row. If the answer to either question 1 or 2, above, is a yes or a maybe, then an “X” is placed in the corresponding cell. At the right of the table is a column that indicates if the row should be considered in the trade space. This condition is met when at least one cell in the row is marked with an “\textbullet”. 

The paragraphs are organized by the columns the broad areas of Space 3.0 and some explanation why and how the application of these technologies and attributes result in at least one “yes” or “maybe” earning that cell an “\textbullet”.


\label{appA} 

\begin{table*}[t]  
\centering
    \begin{tabular}{ 
c||c|c|c|c|c|c||c|
}  
        \toprule 
        Name & \multicolumn{6} {c|} {Space 3.0 Attribute} \\ 
        \midrule 
          & \rotatebox{90}{Servicing, Assembly and Logistics} & \rotatebox{90}{Advanced Communiations} & \rotatebox{90}{Artifical Intelligence/machine Learning} & \rotatebox{90}{Launcher Options} & \rotatebox{90}{Advanced Computing} & \rotatebox{90}{Market} & \rotatebox{90}{Any help?}  \\
        \midrule 
        Closed Architecture &\textbullet  &\textbullet  &  & \textbullet & \textbullet  &\textbullet  &\textbullet   \\
        \midrule
        Technology Antiquity &\textbullet  &\textbullet  &  &  &   &  \textbullet &\textbullet   \\ 
        \midrule
        Risk &\textbullet  &  & \textbullet &  &   &  &\textbullet   \\ 
        \midrule
        Single Launch Mission Architecture &\textbullet  &  &  & \textbullet &   &\textbullet  &\textbullet   \\ 
        \midrule
        Communicatons Rates and Costs &  &\textbullet  & \textbullet &  & \textbullet  &\textbullet  &\textbullet   \\ 
        \midrule
        Operations Cost &\textbullet  &\textbullet  & \textbullet & \textbullet & \textbullet  &\textbullet  &\textbullet   \\
        \midrule
        Long Life: Power &\textbullet  &  & & \textbullet &   &  &\textbullet   \\
        \midrule
        Long Life: Space Craft &\textbullet  &\textbullet  &  & \textbullet &   & &\textbullet   \\
        \midrule
        Cost of Development &\textbullet  &  & \textbullet &  & \textbullet  &\textbullet  &\textbullet   \\
        \midrule
        Cost of Science Instruments &\textbullet  &\textbullet  & \textbullet &  & \textbullet  &  &\textbullet   \\ 
        \midrule
        Annual Funding Limits &\textbullet  &\textbullet  &  &  & \textbullet  & &\textbullet   \\
        \midrule
        Verification (Observatory) &\textbullet  &\textbullet  & \textbullet & \textbullet &   &\textbullet  &\textbullet   \\ 
        \midrule
      
    \end{tabular} 
    \vspace*{5mm} 
     \caption{Summary table for the where the areas that define Space 3.0 can aid in the design of HWO, by reducing schedule span, cost, risk or the by introducing novel capabilities relevant to the mission.}
    \label{tab:my_label} 
\end{table*} 


\subsection{Servicing, Assembly and Logistics}

As presented above, the harbinger of Neo-Space is robotic servicing. The expansion of these services and their commercial development allows for credible for post launch operations science instrument replacement. Also to realize larger apertures, and create greater scientific performance, the ability to service and assemble after launch opens the door to larger apertures than are currently under consideration. The inclusion of assembly as a study may well relieve a major cost and risk driver and one the Tyrannies of the Fairing, that of packaging all the elements of HWO, telescope, instruments and shield in one launch.

Replacement of consumables is also an area of potential benefit. Propellant is the item that first comes to mind with servicing and it should be an element of the trade space. There is another possible consumable replacement that is of special interest to HWO, namely the potential to transfer cryogenic fluids.  It is possible that a baseline or future instrument might require cryogenic temperatures for the instrument. A closed loop cooler exports forces and may not be compatible with an ultra-stable system. The ability to refill a dewar or other cryo reservoir may allow for long lived ultra-low disturbance cryogenic performance.

Servicing may also be of aid in providing a long lived, or more properly a long performing spacecraft function.  The kind of service could be replacing components or functions that have lost performance.
Servicing also has the ability to allow for the program design that permits the program to launch with a completed spacecraft and telescope and a threshold suite of instruments. This type of program architecture could mitigate cost risk instrument development.  A planned later servicing flight would install the instruments that completed development after launch.  This allows the observatory to begin operation at the earliest date. 

\subsection{Advanced communications}

At the very highest level, HWO is a machine in space, at Sun-Earth L2(SE L2), that produces data, in principle a great deal of data.   Getting that product back on the ground, where it can be turned into useful scientific and operational products is usually an expensive endeavor and sacrifices and expense is incurred in crafting design and operation to meet this bottleneck. The projection of the communication rate capability for missions in SE L2 is in the range of 10 to 100 Gb by the 2040s when HWO initial operations are expected.  Furthermore, these rates are expected to grow over time and become less expensive. The reason for this rosy future is that at some level ALL space missions are systems or boxes in space that produce data, if more data can be produced or processed, that system becomes more cost effective and profitable.  So, it is reasonable to expect that the market will demand ever great capabilities here.

It is also near certain that the market may provide commercial communications network for data backhaul. This is a likely application of some of the first LEO mega constellations like Starlink and Kupier. 

The third reason for including the application of these new communications technologies as being of value to HWO involves the instruments. For a Space 2.0 mission, the instrument suite costs are in the range of 20-35 percent of mission costs. \cite{Stahl20},\cite{Feinberg},\cite{CostEffMissons}  Using these proportions and assuming and a second generation of instruments, the fraction of mission costs rises to 33-51\%  of mission cost, likely (and rightly) the most expensive element of the program. Keeping instruments and their modes simple was  key lesson in the development of Spitzer and we would wise to remember and apply it to HWO.\cite{Spitzer}   Given the size and modularity of optical communication systems, it is not unreasonable to imagine a second generation of instruments with their own integral communications systems so as not to be bottlenecked by the HWO system frozen in time at launch!

The argument presented is enough to earn a "\textbullet" in the advanced communications column, but there are still more reasons to study this important technology for inclusion in the HWO architecture, including cost-effective continuous contact with the ground, helping in the integration of HWO into the future time-domain system, including cross-links between observatories in space. The authors, an engineer at that, note that time domain science was highlighted as a key area of investigation in a recent decadal survey.\cite{Decadal} 

\subsection{Artificial Intelligence and Machine Learning }

Artificial intelligence and machine learning (AI/ML) are invading many aspects of modern life and business.  That AI/ML is currently being studied to manage large constellations and spacecraft in general is hardly surprising. The goal of these efforts is to reduce hours operators spend on mundane tasks allowing them to provide more sophisticated trend analysis.  Such trend analysis might be maintenance prediction, similar to commercial vehicles such as cars and planes that report when maintenance is required.  In short,  AI/ML will allow the ground and operations staff be more productive and efficient.  

Additionally, such tools could be used on the ground to aid in possible integration and test simplifications and verification, providing additional opportunities for application of AI/ML to reduce span time and cost.

\subsection{Launcher Options}

In many public statements NASAs astrophysics management regards the emergence of large launchers an enabling technology for HWO.  These large rockets address the tyranny of the fairing directly by greatly reducing mass and volumetric limits on design. These increased margins allow us to consider previously addressable trades, such as trade additional redundancy with servicing for some spacecraft components and larger fuel reserves with servicing costs as first but surely not only examples. Inclusion of large launchers in the design options for HWO eases burdens on the systems designers but does not make them vanish entirely!

The evolution of launchers includes lower cost of access to space. So for the first time it is possible to conceive and execute space flight for the sole purpose of engineering development and model validation.  Much lower costs to space my allow for engineering developments and verifications needed for HWO to be performed directly in space  and not limited development to a lengthy progress through balloon and sounding rockets. \cite{Shkolnik}\cite{Miles}

\subsection{Advanced Computing}

Advanced computing, quantum computing, has the promise of revolutionizing scientific computing. The potential impact on the design of future systems like HWO can only be addressed as hyperbole. However, many researchers world wide are working on this technology and when it arrives it will allow for analysis in greater depth and unprecedented speed of HWO designs at levels of nuance that are not in the current state of the art.

\subsection{Market} 

Perhaps the greatest impact on HWO from the coming Space 3.0 revolution is the emergence of a vibrant market for space technology, missions and operations.  Adapting to this reality might be the hardest thing we have to because it means a cultural or philosophical change in our approach to the problem of HWO, writ large.  There will be cases, where HWO designers of the future will identify a commercial product or service that is close to meeting the needs of HWO and must decide, do we adapt the HWO to use the market provided solution or “do it ourselves”?  Those of use working on the formative stages of this program must begin to adopt these attitudes so they become part of the culture that lets HWO mature at the earliest time and lowest cost. 

\section{Summary and Next Steps}

At the outset of this writing the advancement of the CML for servicing was focused on developing a set of items that might be serviced or replaced, allowing trades to be conducted and moving to CML 3. Upon closer examination this simple list of orbitally replaceable units is not the whole story. We must consider HWO in the engineering milieu it will developed in and we have defined as the emerging neo-space age or Space 3.0. The attributes of this emerging era have been shown to have a significant impact on the design and architecture of HWO to the benefit of the program.  But we must also be wary, striking the right balance of servicing and 'we could do this or that'-ism or HWO will be 'over insured'. Given the finite cost and schedule resources that HWO is likely to face, this 'over-insurance' is an opportunity cost that will be forced on HWO in effect overpaying for a suboptimal design. This opportunity cost must be minimized.  It is recognized that making these kind of decisions will be challenging, describing the process of making these choices is under study and will be reported soon.

We must also remember that the promise of Space 3.0 does not remove the big fundamentals problem of design for HWO, that of achieving ultra-stability. This will no doubt be a great challenge. Space 3.0 offers us more paths to success against this problem and represents a set of invaluable tools to achieve the goal of HWO mission success. The space 3.0 milieu must be part of the architecture study of HWO now, or these advantages will never be included in the HWO design and the promise of the future will be put off to another, later mission. This later flagship will have a launch date will almost surely be in the 2050's under the current development paradigm. 
\\

\textbf{\emph{I don’t want to wait, let’s get the revolution started.}}
\\

{\bf Acknowledgments.} This work was performed under a combination of Northrop Grumman internal funds and the author’s personal time. Special thanks to the subject matter experts from Northrop Grumman who freely gave their time to talk over ideas for this paper, Scott Nuccio, Amber Bauermeister and James Skrinska.

\bibliography{author.bib}

\end{document}